\documentclass[aps,prl,superscriptaddress,twocolumn,showpacs,notitlepage]{revtex4-1}

\usepackage{amsmath}
\usepackage{amsfonts}
\usepackage{amssymb}
\usepackage{hyperref}
\usepackage{graphicx}
\usepackage{color}
\usepackage{xcolor}
\usepackage[mathscr]{euscript}
\usepackage{rotating}
\usepackage{natbib}

\newcommand{\ud}[1]{{#1^{\dagger}}}
\newcommand{\bra}[1]{\left\langle #1\right|}
\newcommand{\ket}[1]{\left| #1\right\rangle}

\newcommand{\rhoel}[1]{\ket{#1}\!\!\bra{#1}}

\newcommand{\mean}[1]{\langle#1\rangle}

\begin{document}
\flushbottom

\title{Criterion for Single Photon Sources}

\author{J.~C. {L\'{o}pez~Carre\~{n}o}}
\affiliation{Departamento
de F\'isica Te\'orica de la Materia Condensada, Universidad
Aut\'onoma de Madrid, 28049 Madrid, Spain}

\author{E. {Zubizarreta Casalengua}} 
\affiliation{Departamento
de F\'isica Te\'orica de la Materia Condensada, Universidad
Aut\'onoma de Madrid, 28049 Madrid, Spain}

\author{E.~del~Valle}
\affiliation{Departamento de F\'isica Te\'orica de la Materia
Condensada, Universidad Aut\'onoma de Madrid, 28049 Madrid,
Spain}

\author{F.~P.~Laussy}
\affiliation{Russian Quantum Center, Novaya 100,
  143025 Skolkovo, Moscow Region, Russia}
\affiliation{Departamento de
  F\'isica Te\'orica de la Materia Condensada, Universidad Aut\'onoma
  de Madrid, 28049 Madrid, Spain}

\begin{abstract}
  We propose a criterion to classify and compare sources of single
  photons: the $p$-norm of their $k$th-order photon
  correlations. While ideally one should use the uniform norm
  with~$p\rightarrow\infty$, in practice, the 3-norm already allows a
  meaningful comparison and shows the serious limitations of the usual
  criterion based on the second order correlation~$g^{(2)}$ only. We
  apply this new criterion to a large family of popular single photon
  sources and show that some of them do not, in fact, qualify for this
  function. The best sources from our selections consist of 
  a cascade of two-level systems, improving on the otherwise standard
  low-power resonant excitation of a single two-level system.
\end{abstract}

\date{\today}

\maketitle 

Single photon sources (SPS)~\cite{aharonovich16a} constitute a basic
component of the future quantum technologies~\cite{obrien09a} and
their laboratory implementation is therefore actively pursued by
several groups that compete for the delivery of the best emitters with
the required
characteristics~\cite{kuhlmann15a,somaschi16a,ding16a,wang16a,kim16b}. The
desiderata for a good SPS cover a large range of both fundamental
(antibunching, indistinguishability, etc.) and applied (efficiency,
price, etc.)  interest~\cite{dada16a, mueller16a, eisaman11a}. Since
the technology is yet largely within a development phase, the most
prized characteristics are still those of a fundamental character, and
there is a need to compare different emitters from various platforms
and operating in different conditions. One of the important requisites
is the suppression of multiple-photon events, which is usually
characterized by the zero-delay second-order correlator function
$g^{(2)}=\mean{\ud{\varsigma}^2\varsigma^2}/\mean{\ud{\varsigma}\varsigma}^2$
that quantifies the correlations between two quanta from a field with
annihilation operator~$\varsigma$~\cite{paul82a}.  It has long been
known that $g^{(2)}<1$, linked to sub-Poissonian fluctuations, is a
proof of the non-classical character of the field and is commonly used
as a criterion for the quantum character of a state.  Such a condition
accommodates as well, however, a weak-class of quantum states that
consists of convex mixtures of Gaussian states~\cite{filip11a}, while
SPS are pursued for providing stronger quantum resources that allow,
for instance, the reconstruction of any other state.  A single photon,
being a very particular case, requires a dedicated way to qualify it.
The criterion is simple enough but is not practical, it reads:
$g^{(2)}=0$. However, a mathematical zero is not a good criterion for
a physical observable, that is prone to error and
approximations. Aiming instead for this quantity to be as close to
zero as possible would be the natural remedy but is in fact not
adequate, as we discuss in this Letter. Consequently, we propose a
fitting substitute and use it to establish a first comparison between
different mechanisms that power SPS.

The fundamental emitter of single photons is the two-level system
(2LS) with annihilation operator~$\sigma$. By construction, the 2LS
can only support one excitation, so it can only emit one photon at a
time. The $g^{(2)}$ of its emitted light is, theoretically, exactly
zero. This is one example of a mathematical zero since the
cancellation of~$g^{(2)}$ when~$\varsigma=\sigma$ occurs at the level
of the nilpotent operator: $\sigma^2=0$. This is therefore realized
for any excitation of the system. On the detection side, however, the
situation is not equally ideal. In fact, already the strong quantum
character of the state is affected by the contribution of the vacuum.
This has been recently shown~\cite{filip11a} through an upper bound
for the probability~$p(1)$ of the one-photon component for the class
of convex mixtures of Gaussian states: no such state can ever be found
with this probability overcoming
$3\sqrt{3}/(4e)\approx 0.477889$~\cite{filip11a}. This provides a
criterion for the strong quantum character of the state. For a 2LS
with decay rate~$\gamma_\sigma$ that is excited incoherently at the
rate~$P_\sigma$, this is achieved
when~$P_\sigma/\gamma_\sigma>p(1)/(1-p(1))\approx0.915303$. There is
therefore a threshold in pumping to meet this criterion. Higher
pumping of course satisfy it even more and in the limit of infinite
pumping, the incoherent excitation of a 2LS maintains the pure Fock
state~$\ket{1}$ in the steady state, since the system is forced to
remain excited. This is the most quantum state one can get with
respect to a single-photon objective. Strong pumping is not, however,
a good approach to SPS, since it spoils another important quality:
their indistinguishability. With pumping comes power broadening and
photons are emitted with increasingly fluctuating energies. The Fock
state~$\ket{1}$ emits photons one at a time but at all the
frequencies!

The photon frequencies are in fact a crucial component when it comes
to suppressing multiple-photon emission. In practice, there are
numerous reasons to get accidental coincidences, but one is of a
fundamental character, and is related to the tail events in the power
spectrum (the energy distribution of the emitted photons). This is
fundamental because the power spectrum for spontaneous emission of a
2LS is Lorentzian which means that it fluctuates to all orders and it
is therefore fundamentally impossible to detect all the photons (since
it is impossible to detect at all the frequencies). In contrast, a
distribution with a standard deviation would allow, for all practical
purposes, to detect all the photons given a wide enough filter (say
five standard deviations). Interposing a filter between the emitter
and the detector (this also describes the detector's finite bandwidth)
affects the statistics of the detected photons~\cite{delvalle12a}.
Thus, although the 2LS can host only one excitation, its filtering
results in nonzero probabilities~$p(n)$ to detect~$n\ge2$ photons.
There are various ways to understand this result: the filter
introduces a time uncertainty that makes it possible to pile-up
consecutive photons.  From a quantum mechanical viewpoint, the 2LS
emits a continuous stream of photons at all the possible frequencies,
which amplitudes interfere destructively to result in a single photon
upon detection. Trimming the tails spoils the destructive interference
and allows more than one photon to reach the detector.

\begin{figure}[t!]
  \centering
   \includegraphics[width=0.85\linewidth]{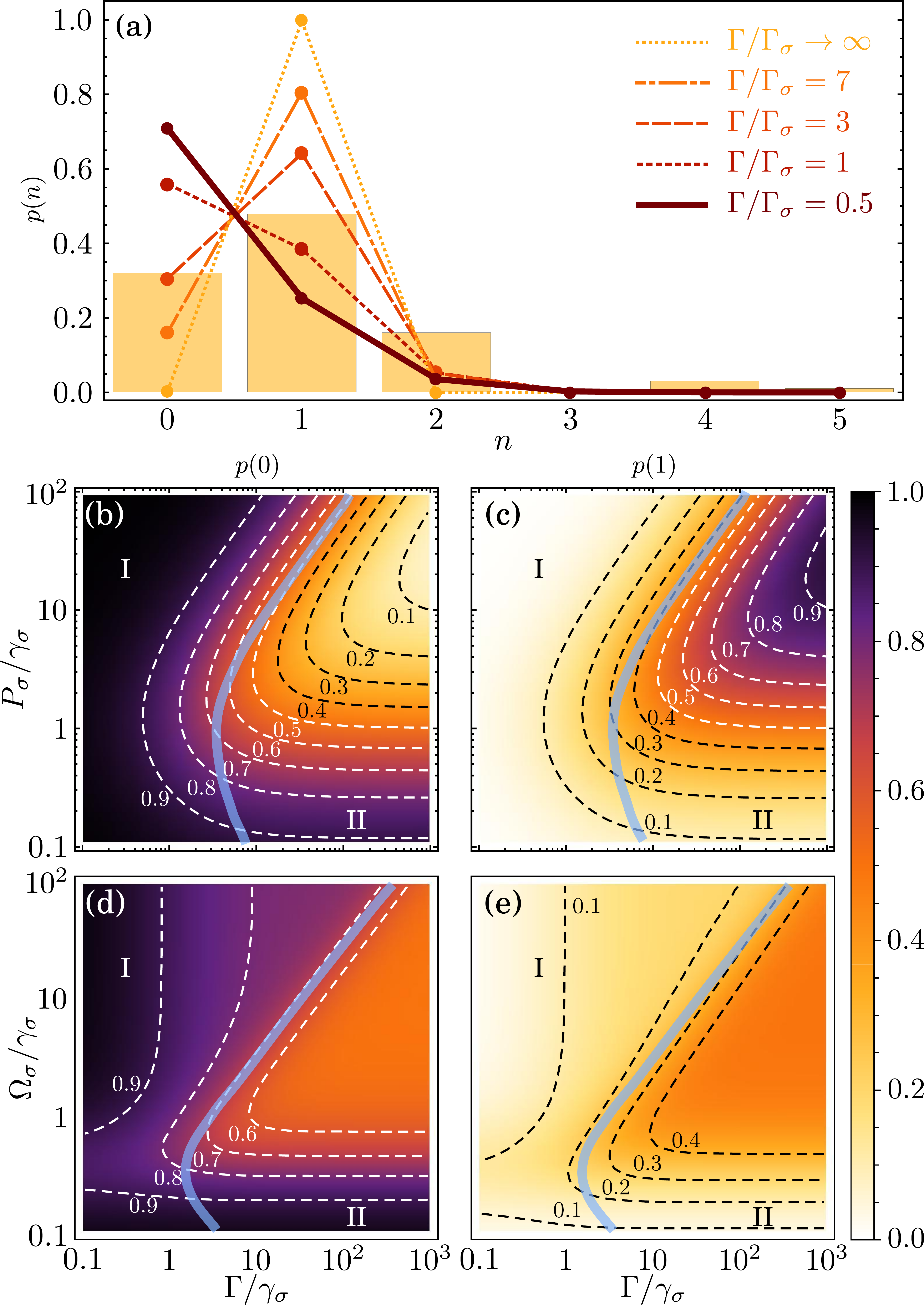}
   \caption{(Color online). (a)~Probability distribution of the
     Gaussian state maximizing the probability of one
     excitation~\cite{filip11a}~(bars), and distributions of a
     filtered 2LS incoherently pumped, with
     $P_\sigma/\gamma_\sigma=100$, for several filter linewidths.  For
     vanishing filter linewidths, $\Gamma\rightarrow 0$, the
     distribution is thermal (dark solid thick line). In the opposite
     limit, $\Gamma\rightarrow \infty$, the unfiltered case goes to
     $\ket{1}$ for high enough pumping (light dotted line). Panels~(b)
     and~(d) show the probability of the vacuum and panels~(c) and~(e)
     of having one excitation, for incoherent (b~\&~c) and coherent
     (d~\&~e) excitation.  The blue thick contour divides the states
     that exist as a convex mixture of Gaussian states (I) from those that
     exhibit a strong quantum character (II)~\cite{jezek11a}.  }
  \label{fig:MonSep19223347CEST2016}
\end{figure}

The distributions for an incoherently excited 2LS that is detected in
various frequency windows is shown as joined dots in
Fig.~\ref{fig:MonSep19223347CEST2016}(a). Here the pumping was very
large to keep the system in its excited state.  This is obtained
through a master equation for the incoherent excitation (at a
rate~$P_\sigma$) of a 2LS with free Hamiltonian
$H_\sigma=\omega_\sigma\ud{\sigma}\sigma$ and decay rate
$\gamma_\sigma$ (we take $\hbar=1$ along the text):
\begin{equation}
  \label{eq:TueSep20193107CEST2016}
  \partial_t \rho = i[\rho, H_\sigma] +
  \frac{\gamma_\sigma}{2}\mathcal{L}_\sigma\rho+
  \frac{P_\sigma}{2}\mathcal{L}_\ud{\sigma}\rho\,,
\end{equation}
where $\mathcal{L}_c\rho = (2c\rho \ud{c}-\ud{c}c\rho-\rho\ud{c}c)$.
The filtering, or the finite temporal and spectral resolution of the
detectors, are taken into account self-consistently through the theory
of frequency-filtered and time-resolved $N$-photon
correlations~\cite{delvalle12a}, where a sensor with spectral width
$\Gamma$ (and hence with temporal resolution~$1/\Gamma$) is included
into the dynamics of the system in the limit of its vanishing coupling
to the SPS, so that it only ``senses'' correlations without perturbing
the system.  For an incoherently driven 2LS, the $n$th-order filtered
correlation function~$g_{\Gamma}^{(n)}$ can be thus obtained recursively
using the relation:
\begin{equation}
  \label{eq:marsep13180042CEST2016}
  g_\Gamma^{(n)}=g_{\Gamma}^{(n-1)}\frac{n\Gamma_\sigma
    (\Gamma_\sigma+\Gamma)}{[\Gamma_\sigma+(n-1)\Gamma]\left [
      \Gamma_\sigma+(2n-1)\Gamma\right]}\, ,
\end{equation}
where $g_\Gamma^{(1)}= 1$ and
$\Gamma_\sigma\equiv\gamma_\sigma+P_\sigma$. Since the filtered
$n$-photon correlations are nonzero, the effective density matrix that
describes the state of the 2LS is clearly no longer constrained to one
excitation. The probability $p(n)$ to find in the Fock state $\ket{n}$
the effective emitter that corresponds to a filtered 2LS can be
recovered through the bijective mapping provided in
Ref.~[\onlinecite{arXiv_zubizarreta16a}]:
\begin{equation}
  \label{eq:MonSep19222001CEST2016}
  p(n) = \sum_{k\geq n}\frac{(-1)^{k+n}}{n!\,(k-n)!}G_{\Gamma}^{(k)}\,,
\end{equation}
where $G_\Gamma^{(k)}=\langle\ud{\sigma}\sigma\rangle^k_\Gamma
g_\Gamma^{(k)}$ is the $k$th order unnormalized correlator, and
$\mean{\ud{\sigma}\sigma}_\Gamma=P_\sigma
\Gamma/(\Gamma_\sigma\Gamma+\Gamma^2_\sigma)$ is the population that
passes through a Lorentzian bandpass filter. Combining
Eqs.~(\ref{eq:marsep13180042CEST2016}-\ref{eq:MonSep19222001CEST2016}),
we obtain the distribution of the filtered SPS in closed form:
\begin{equation}
  \label{eq:marsep27161356CEST2016}
  p(n) = \mathcal{C}_n\times
  {}_1\!F_2\left(n+1;\frac{2n+1}{2}+\frac{\Gamma_\sigma}{\Gamma},
    n+\frac{\Gamma_\sigma}{\Gamma};
    -\frac{P_\sigma} {2\Gamma}
  \right)\,,
\end{equation}
where ${}_1\!F_2$~is the generalized hypergeometric function
and~$\mathcal{C}_n$ are coefficients that satisfy:
\begin{equation*}
  \label{eq:marsep27162136CEST2016}
  \frac{\mathcal{C}_n}{\mathcal{C}_{n-1}}=\frac{P_\sigma\Gamma}{
    [\Gamma_\sigma+(n-1)\Gamma][\Gamma_\sigma+(2n-1)\Gamma]}\quad {\rm
    with}\quad \mathcal{C}_0=1\,.
\end{equation*}
These expressions, displayed in
Fig.~\ref{fig:MonSep19223347CEST2016}(a), show that filtering (or,
equivalently, using a low-resolution detector) spoils both~$g^{(2)}$
and the strong quantum character of the source, as it brings it to
break the Filip \& Mi\v{s}ta aforementioned criterion~\cite{filip11a}
already for a filter (or detection) bandwidth~$\Gamma$ commensurate to
the power broadened linewidth~$\Gamma_\sigma$ (we also show as a bar
plot the probability distribution for the Gaussian state that
maximizes the probability of one excitation~$p(1)$).  Narrower filters
completely thermalize the signal, with $g^{(n)}\rightarrow n!$
when~$\Gamma/\Gamma_\sigma\rightarrow0$.
Figures~\ref{fig:MonSep19223347CEST2016}(b--c) show the probability
for the vacuum~$p(0)$ and for one-excitation~$p(1)$ as a function of
the incoherent pumping rate and filtering of the 2LS.  Taken together,
these two quantities provide a stricter criterion to separate weak
classes of quantum states from strong ones~\cite{jezek11a}. Namely,
the parameter-space that lies on the right of the thick blue line
realizes steady state of the 2LS that cannot be represented as a
convex mixture of Gaussian states, i.e., are the quantum non-Gaussian
states sought for quantum applications. The processes shaping such
states require high-order quantum nonlinearities.  This confirms that
to maintain the quantum character of the source, one requires a
minimum filter linewidth. For a pumping rate commensurable with the
decay rate, such a minimum is roughly twice the intrinsic linewidth of
the 2LS. As could be expected, for higher pumping rates the
power-broadening implies that a correspondingly larger filter is
required. However, at vanishing pumping, this time unexpectedly, one
also needs increasing filtering, showing how the more scarcely the 2LS
is excited, the less quantum it is.  This comes from the deleterious
contribution of the vacuum.

Since the tail events are the main responsible for spoiling the SPS,
an emitter whose power spectrum is not Lorentzian but has faster
decaying tails would allow to collect more of the photons in a reduced
spectral window and thus better preserve its quantum features. The
simplest way to achieve that is to turn to coherent driving, in which
case Heitler processes of coherent absorption and emissions trim the
Lorentzian tails to yield instead a Student $t_2$ distribution, which
still has fat tails but that decrease faster than a
Lorentzian~\cite{arXiv_lopezcarreno16b}. The same idea can be further
enforced by turning to quantum rather than classical excitation, in a
cascaded scheme of several 2LS in a
chain~\cite{arXiv_lopezcarreno16b}.  A drawback of coherent driving is
that it forbids total inversion of the 2LS, which, due to stimulated
emission, is at most half-excited. Therefore, while $g^{(2)}$ is
indeed smaller, the quantum state itself deviates less from classical
mixtures than under incoherent excitation. This is shown in
Figs.~\ref{fig:MonSep19223347CEST2016}(d--e) that are the counterpart
of the above row but for coherent excitation. The probability
saturates at $p(1)=1/2$, which is a considerable impediment when
trying to demonstrate the genuine quantum character of a source (for
instance through the measurement of a negative Wigner function).

From our discussion so far, it is clear that the quantum non-Gaussian
character of the state, although of great fundamental interest, is not
the best practical way to assess the quality of a SPS. The widely used
criterion that rely on $g^{(2)}$ is more adequate and indeed sets the
standard. It presents, however, some subtleties and is frequently
mishandled, for instance through a popular criterion $g^{(2)}<1/2$
that supposedly guarantees the single-photon or single-emitter
character of the
source~\cite{michler00a,verma11a,dimartino12a,lukishova12a,palaciosberraquero16a}.
We have shown in an earlier work~\cite{arXiv_lopezcarreno16a,
  arXiv_zubizarreta16a}, however, that the Hilbert space contains a
myriad of states with population larger than one that satisfy the
condition~$g^{(2)}<1/2$. The exact criterion remains $g^{(2)}=0$. It
is also invalid, however, to aim for $g^{(2)}\rightarrow 0$, as states
with arbitrarily small but nonzero $g^{(2)}$, but featuring a huge
$g^{(n)}$ for~$n\ge3$, are possible. The state $\rho =
(297001/300000)\rhoel{0}+(1999/200000)\rhoel{1}+(1/600000)\rhoel{3}$
for instance has $g^{(2)}=1/10$ and $g^{(3)}=10$, so stopping at the
second-order correlation does not well characterize a SPS. Clearly, in
the most general setting, a SPS has to exclude all fanciful designs,
possibly deviously conceived fake SPS, that would give the illusion of
good antibunching through standard~$g^{(2)}$ measurements, while
opening photon-number splitting backdoors for hacking cryptographic
protocols~\cite{brassard00a} by encoding the information in~$N$ photon
states not detectable at the two-photon level.

Here, we propose a novel criterion to well identify and compare
usefully SPS, by using all the correlation functions. Namely, we 
propose the~$N$-norm of the vector of correlation functions:
\begin{equation}
  \label{eq:MonSep19224648CEST2016}
 \| ( g^{(k)} )\|_{N} \equiv
  \sqrt[N]{\sum_{k=2}^{N+1}{g^{(k)}}^N}\,,
\end{equation}
that quantifies the deviation between the ideal SPS, which is
perfectly antibunched to all orders, and an actual source.  The
criterion is then that~$\|(g^{(k)})\|_{N}$ should be as small as
possible (not simply the~$g^{(2)}$, which is the 1-norm) for $N$ as
high as technically feasible. In the limit $N\rightarrow\infty$, the
quantity becomes $\|( g^{(k)} )\|_\infty=\max(g^{(k)})$ and measures
the failure of the SPS by the order at which it less suppresses
multiple-photon emission.

\begin{figure}[tb!]
  \includegraphics[width=0.95\linewidth]{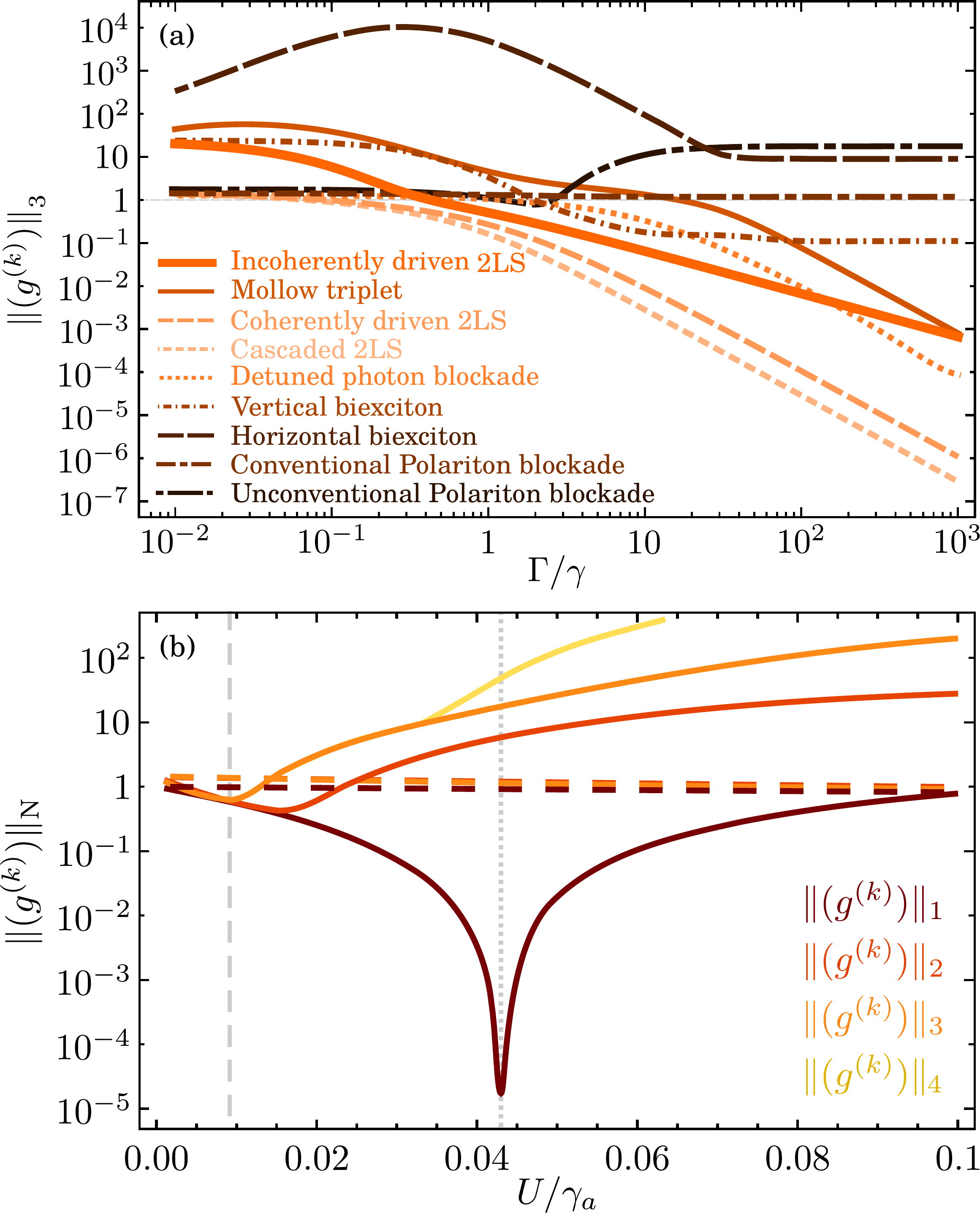}
  \caption{(Color online). Criterion for Single Photon
    Sources. (a)~3-norm for several sources of light as a function of
    their spectral filtering. The incoherently driven 2LS (in the
    limit of $P_\sigma\ll \gamma_\sigma$) is set as the benchmark for
    SPS. It is always surpassed by the coherently driven 2LS and the
    cascaded 2LS as long as $\Omega_\sigma\ll\gamma_\sigma$. The
    emission of the biexciton (with $\chi/\gamma_\sigma =40$ and
    $\Omega_\sigma/\gamma_\sigma=10$), on the other hand, fails to
    overcome the benchmark in either polarization. The blockade
    mechanisms provide better antibunching in the limit of vanishing
    $\Gamma$, but fail as SPS in the opposite limit. Here~$\gamma$ is
    meant as either~$\gamma_\sigma$ or~$\gamma_a$ depending on the
    system. (b)~Comparison of the $\|(g^{(k)})\|_N$ for the
    conventional~(dashed) and unconventional~(solid) polariton
    blockade (Parameters as in Ref.~\cite{lemonde14a}, i.e.,
    $\gamma_a=\gamma_b$, $\omega/\gamma_a=0.275$ and $g/\gamma_a=3$.
    The filtered $3$-norm in~(a) is for $U/\gamma_a=0.0425$
    that minimizes~$g^{(2)}$.}
  \label{fig:MonSep19223913CEST2016}
\end{figure}
In Fig.~\ref{fig:MonSep19223913CEST2016}(a), we show the result for
several popular sources, limiting to the $3$-norm (i.e., measuring up
to~$g^{(4)}$), which is enough for the cases we have chosen (that do
not engineer scenarios where the SPS character breaks for a large
number of photons). We show the results as a function of the filtering
spectral resolution, that, as we have already discussed, is a key parameter
for photon correlations.  The incoherently driven 2LS sets a useful
reference. As the filtering window is enlarged, our
measure~(\ref{eq:MonSep19224648CEST2016}) indeed decreases.  The
coherently driven 2LS, in comparison, decreases much faster, and this
confirms the known fact that resonant excitation provides better
antibunching than the non-resonant case. This is the case at low
driving, since at high driving, the system enters the Mollow triplet
regime~\cite{mollow69a}, in which case the statistics of the emission
varies strongly with the frequency~\cite{gonzaleztudela13a}. In
particular, the emission at frequencies between the triplet's central
and lateral peaks is largely bunched, as opposed to the resonant
emission which remains antibunched.  Even though performing as a SPS
when filtering over the entire structure, this affects considerably
the performances of the 2LS.  The cascading scheme that we have
recently proposed~\cite{arXiv_lopezcarreno16b} and already discussed
above is confirmed as a better SPS than under classical resonant
driving, although the slope is the same as for the conventional
case. It is in fact the best SPS of our selection.

A value of our criterion resides in its identification of sources with
a strong antibunching (small~$g^{(2)}$) but that fail in other
respects. This the case for instance of a biexciton, i.e., a molecule
that consists of a bound pair of excitons (an electron-hole pair). It
supports two excitons with opposite third component of the angular
momentum, namely~$+1$ and~$-1$, which are represented by the
annihilation operators~$\sigma_\uparrow$ and~$\sigma_\downarrow$,
respectively. The linearly polarized states of the biexciton system,
namely~$\sigma_{\mathrm{H,V}}=(\sigma_{\uparrow} \mp
\sigma_{\downarrow})/\sqrt{2}$,
describe transitions from the exciton to the ground state emitting a
vertically and horizontally polarized photon, respectively. To test
the biexciton as a SPS, we drive it coherently with a vertically
polarized laser of amplitude~$\Omega_\sigma/\!\sqrt{\gamma_\sigma}$
and in resonance with the exciton transition.  The Hamiltonian
describing the biexciton is given by
$H_\mathrm{B}= \omega_{\sigma}
(\ud{\sigma_{\uparrow}}\sigma_{\uparrow} +
\ud{\sigma_{\downarrow}}\sigma_{\downarrow}) - \chi
(\ud{\sigma_{\uparrow}}\sigma_{\uparrow}
\ud{\sigma_{\downarrow}}\sigma_{\downarrow}) + \Omega_\sigma
(\ud{\sigma_{\rm V}}e^{i\omega_\sigma t}+ \sigma_{\rm
  V}e^{-i\omega_\sigma t})$
where~$\omega_\sigma$ is the bare exciton energy (we are considering
resonant excitons) and $\chi$ is the bonding energy, so that the
biexciton energy is~$2\omega_\sigma-\chi$. The biexciton is described
by the master equation given in Eq.~(\ref{eq:TueSep20193107CEST2016})
replacing $H_\sigma$ by~$H_\mathrm{B}$, dropping the term related to
the incoherent driving~($\mathcal{L}_{\ud{\sigma}}\rho$), and letting
the excitons decay with rate~$\gamma_\sigma$ for both
polarizations. Figure~\ref{fig:MonSep19223913CEST2016}(a) shows that
the 3-norm for the emission of the biexciton saturates above the
measure set by the incoherently driven 2LS. This means that, as
expected, the biexciton is not a good SPS. However,
Fig.~\ref{fig:MonSep19223913CEST2016}(a) also shows that the asymmetry
in the driving field favors the vertically polarized emission, which
performs better than the horizontally polarized emission. In fact, for
a large region of values for $\Gamma/\gamma$, the vertically polarized
emission of the biexciton is a better SPS than the central peak of the
Mollow triplet.

Another striking case is the so-called ``polariton blockade'' (also
known as ``photon blockade''), in both its
conventional~\cite{verger06a} and
unconventional~\cite{liew10a,bamba11a,flayac13a} form.  In the former
case, antibunching follows from self-interactions (polaritons in this
context are essentially strongly interacting photons), while in the
latter case, a much stronger antibunching is obtained at vanishing
pumping from an interference. Calling $a$ and~$b$ the annihilation
operators of two polariton modes, the Hamiltonian reads
$H_\mathrm{P} = \omega(\ud{a}a +\ud{b}b) + g(\ud{a}b+\ud{b}a) +
U(\ud{a}\ud{a}aa+\ud{b}\ud{b}bb)+\Omega_a(\ud{a}e^{i\omega_{\rm L}t}+a
e^{-i\omega_{\rm L}t})$
where~$\omega$ is the polariton energy, $g$~is the coherent coupling
between polaritons (e.g., by tunneling), $U$~is the strength of
polariton interaction, $\gamma_a$ is the polariton decay rate
and~$\Omega_a/\!\sqrt{\gamma_a}$ is the amplitude of the laser driving
one of the polaritons. This Hamiltonian covers both types of blockades
depending on the frequency of resonant excitation and the coherent
coupling.  Figure~\ref{fig:MonSep19223913CEST2016}(a) shows that the
conventional polariton blockade (obtained by setting $g=0$ in
$H_\mathrm{P}$ to keep only polaritons~$a$) behaves differently from
the other SPS, as it remains roughly constant as more light is
collected.  The situation is even more serious for the unconventional
polariton blockade that, contrary to a popular belief that made its
manifestation actively sought after, performs rather poorly as a SPS,
in particular in the configuration where its~$g^{(2)}$ is very small.
Indeed, despite its excellent antibunched $g^{(2)}$ emission, it also
exhibits a superbunched $g^{(k)}$ for $k\geq3$, and therefore fails to
behave as a SPS for most of the values of $U/\gamma_a$, including that
where~$g^{(2)}\ll1$.  Such a discrimination between the two types of
blockades is shown in Fig.~\ref{fig:MonSep19223913CEST2016}(b),
comparing all~$\|(g^{(k)})\|_p$ for~$p=1,\ldots,4$ and this time as a
function of the interaction strength (normalized to the
broadening)~$U/\gamma_a$. The dashed lines correspond to the
conventional blockade and show an essentially converged result
regardless of~$p$, indicating that the conventional blockade is a SPS
that gets increasingly (albeit slowly) better with increasing
nonlinearity. On the other hand, while the unconventional blockade is
much better as perceived through the 1-norm (the
conventional~$g^{(2)}$, this is the result hailed in the literature),
it is quickly spoiled at the 2-norm level, particularly at the value
of $U/\gamma_a$ that minimizes $g^{(2)}$. It remains a better SPS
than its conventional blockade counterpart on the range shown here
(dashed line), but in a different configuration than the one that is
aimed for (dotted line) and would be overtaken by the conventional
blockade for stronger nonlinearities.

With such a criterion in hand, one can devise configurations that
combine the assets of various approaches.  For instance, the bare 2LS
presents the shortcoming of emitting in a large solid angle and
placing it in a cavity brings several advantages, such as increasing
its emission rate by Purcell enhancement, but also for practical
purposes, in collecting and directing the light in a focused output
beam. One can ask whether this has a cost in other aspects and weakens
the supression of multiple-photon emission as compared to bare SPS. We
find with Eq.~(\ref{eq:MonSep19224648CEST2016}) that in the limit of
weak-coupling and fast decay rate of the cavity, the impact is
negligible, which is a welcome result as this configuration is
widespread.  More involved designs can optimize the combination of
these aspects in other configurations, for instance the detuned photon
blockade~\cite{muller15a} proves to remain an excellent
SPS within a cavity, making such an implementation a serious contender
for future applications.

In summary, we studied the quantum non-Gaussian character of a 2LS and
how filtering and/or finite-detector resolution is affecting its
quality as a SPS. We proposed a new criterion to quantify its
performance: the $N$-norm of the photon correlation functions.  For
practical purposes, it is convenient and often enough to limit to the
2-norm or 3-norm. On the other hand, we showed that~$g^{(2)}$ alone
(the 1-norm), which is the widespread standard, is not sufficient and
can lead to erroneous assessments.  We illustrated the criterion with
a variety of mechanisms to implement SPS and found that the
yet-to-be-implemented cascaded SPS provide the best realization of
single-photon emitters while resonance fluorescence at low driving are
the best already existing SPS.  We also showed how other sources that
appear to be good SPS do actually fail in a stricter sense that could
jeopardize quantum information protocols.

The POLAFLOW ERC project No.~308136, and the Spanish MINECO contract
FIS2015-64951-R (CLAQUE) are acknowledged.

\bibliographystyle{naturemag}
\bibliography{sci,arXiv,books}

\end{document}